\documentclass[journal]{IEEEtran}

\usepackage{comment}

\usepackage{cite}
\usepackage{amsmath,amssymb,amsfonts}
\usepackage{mathabx}  
\usepackage{algorithmic}
\usepackage{graphicx}
\usepackage{textcomp}
\usepackage{accents}

\usepackage{cite}
\usepackage{amsmath,amssymb,amsfonts}
\usepackage{algorithmic}
\usepackage{graphicx}
\usepackage{epstopdf}
\usepackage{subfigure}
 \graphicspath{{\figures}}
\usepackage{textcomp}
\usepackage{xcolor}
\usepackage{verbatim}
\usepackage{makecell}
\usepackage{booktabs}
\usepackage{CJK}
\usepackage{multirow}
\usepackage{url}

\usepackage{array}

\usepackage{algorithm}
\usepackage{algorithmic}

\makeatletter
\newcommand{\removelatexerror}{\let\@latex@error\@gobble}
\makeatother

\begin{document}

\begin{CJK}{UTF8}{gbsn}

\title{ 
Multimodal Semantic Communication for \\ Generative Audio-Driven Video Conferencing

}
\author{Haonan Tong, Haopeng Li, Hongyang Du, \textit{Student Member, IEEE,}
Zhaohui Yang, \textit{Member, IEEE}, \\
Changchuan Yin, \textit{Senior Member, IEEE}, 
and Dusit Niyato, \textit{Fellow, IEEE}. 
\vspace{-0.99cm} 

\thanks{
H. Tong, H Li, and C. Yin are with the Beijing Key Laboratory of Network System Architecture and Convergence, and also with the Beijing Advanced Information Network Laboratory, Beijing University of Posts and Telecommunications, Beijing, 100876 China, Emails: hntong@bupt.edu.cn, hpli@bupt.edu.cn,  ccyin@bupt.edu.cn.


H. Du 
is with the Department of Electrical and Electronic Engineering, University of Hong Kong, Pok Fu Lam, Hong Kong SAR, China.
Email: duhy@eee.hku.hk.

Z. Yang is with the College of Information Science and Electronic Engineering, Zhejiang University, Hangzhou, Zhejiang 310027, China,
Email: yang\_zhaohui@zju.edu.cn.

 D. Niyato  is 
with the 
College of Computing and Data Science 
at Nanyang Technological University (NTU), Singapore, Email: dniyato@ntu.edu.sg.

}



 } 
\maketitle

\begin{abstract}

This paper studies an efficient multimodal data communication scheme for video conferencing.
In our considered system, a speaker gives a talk to the audiences, with talking head video and audio being transmitted.
Since the speaker does not frequently change posture and high-fidelity transmission of audio (speech and music) is required, redundant visual video data exists and can be removed by generating the video from the audio.
To this end, we propose a wave-to-video (Wav2Vid) system, an efficient video transmission framework that reduces transmitted data by generating talking head video from audio.
In particular, full-duration audio and short-duration video data are synchronously transmitted through a wireless channel, with neural networks~(NNs) extracting and encoding audio and video semantics. 
The receiver then combines the decoded audio and video data, as well as uses a generative adversarial network~(GAN) based model to generate the lip movement videos of the speaker.
Simulation results show that the proposed Wav2Vid system can reduce the amount of transmitted data by up to 83\% while maintaining the perceptual quality of the generated conferencing video.

\end{abstract}

\begin{figure*}[t]
    \centering
    \vspace{-0.1cm}
\includegraphics[width=0.95\linewidth,
    ]{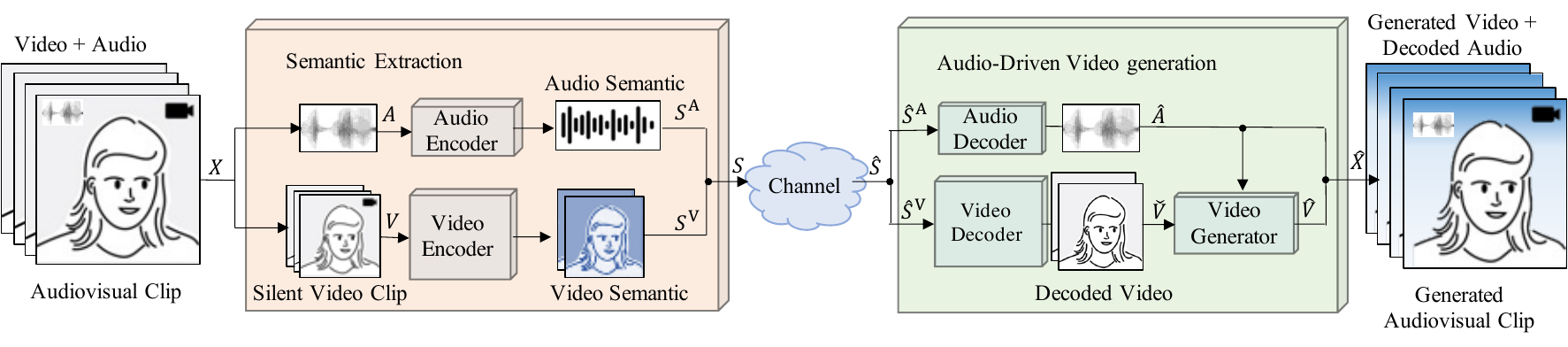}
    \setlength{\abovecaptionskip}{-0.3cm}
    \caption{The architecture of Wav2Vid enabled video conferencing.
    }\label{fig:architecture}
    \vspace{-0.5cm}
\end{figure*}

\vspace{-0.3cm}
\section{Introduction}

Interactive services have rapidly developed in wireless network systems, leading to a significant increase in the volume of video data transmission and has become the main traffic in the network system~\cite{Cisco_report}. 
However, interactive video data imposes substantial communication overloads, while limited communication resources impose constraints on the implementation of interactive applications. 
{\color{black}
Semantic communication on visual data (image and video, etc.), which benefits from joint source-channel coding~(JSCC)~\cite{CDDM}, can reduce the volume of data transmission by extracting visual data semantics according to the communication task, thus improving interactive quality with limited resources.
}



Recently, several works~\cite{Deep_V_SC_ZP, jinshi_v_vonf,li2024video} have studied efficient video communication methods with semantic extraction.
The work in~\cite{Deep_V_SC_ZP} proposed a neural network~(NN) based video semantic communication scheme, which is robust against wireless channel impairments, especially under low signal to noise ratio (SNR) conditions.
Furthermore in~\cite{jinshi_v_vonf}, the authors proposed semantic based video conferencing communication, in which human facial key-points are extracted and efficiently transmitted for portrait reconstruction at the receiver.
The work in~\cite{li2024video} extracted the major objects in videos and reduced transmission overloads by not transmitting the redundant visual parts.
However, the compression ratios of these methods~\cite{Deep_V_SC_ZP, jinshi_v_vonf,li2024video} are limited in video conferencing, because they lack the extraction of redundancy among different modal data, since the visual information in talking head videos in conferencing is closely related to the speaker's speech. 

The prior works in~\cite{Luo_WC,Xing_WCL,JSAC_mit} have focused on multimodal semantic communication.
The work in~\cite{Luo_WC} studied multimodal multiuser semantic communication in a channel-level information fusion to improve the spectrum efficiency.
In~\cite{Xing_WCL}, the authors proposed multimodal semantic communication based on knowledge graphs for visual question answering~(VQA), which improved the vehicle driving efficiency and safety by text clarifying the communication task and thus reduced redundant transmission.
However, the methods in~\cite{Luo_WC, Xing_WCL}, which fused multimodal data to provide supplementary information for enhancing task completion, did not aligned the semantics of the multimodal data and therefore cannot generate cross-modality data.
For aligned semantics in semantic communication, the method in~\cite{JSAC_mit} transmitted text data to enable video conferencing using short-duration text driving videos to generate long-duration videos.
However, since audiences in video conferencing mainly focus on audio information, the text-driven method in~\cite{JSAC_mit} cannot meet the conferencing requirements because it loses key audio information such as the speaker's emotion and emphasis.



To this end, we aim to study multimodal semantic communication for video conference using high-fidelity audio data to generate the speaker's video data.
The main contributions of this paper are summarized as follows.

1)  We propose a multimodal semantic communication-enabled wave-to-video~(Wav2Vid) system for video conferencing. 
The Wav2Vid system transmits the semantics of short-duration video clips and unabridged audio. 
Long-duration visual videos are then generated with these short-duration video clips and full-duration audio.


2) We design the audio codec, video codec, and video generator in Wav2Vid.
The audio codec extracts the audio waveform features, which are coded to mitigate wireless channel impairments.
The video codec extracts and encodes the video spatio-temporal context information.
The video generator extracts relations between video and audio and generates the speaker's lip movements synchronized to the audio.
The codecs use pre-trained models and are fine-tuned to learn the wireless channel character.

Simulation results show that the proposed  Wav2Vid system can generate vivid talking head videos, and 
reduce the transmitted data amount by up to 83\% while maintaining high video perceptual quality.
The remainder of this paper is organized as follows. 
The system model is formulated in Section II. 
The designed semantic codecs are introduced in Section III.
In Section IV, simulation results are presented and discussed.
The conclusions are drawn in Section V.




\color{black}
\section{System Model}

We consider a video conference scenario in which one speaker gives a talk facing a camera in a static environment, as shown in Fig.~\ref{fig:architecture}. 
In this scenario, the speaker's device must send
audiovisual video clips $\boldsymbol X$ to the audience.
Note that, the visual information in $\boldsymbol X$ has redundancy during the video conference, i.e., the main changes in videos are lip movements while the other parts may stay still (background) or change periodically (i.e., eye blink) if the speaker does not have significant posture change~\cite{Wav2Lip}.
To this end, in the Wav2Vid system, only effective video clips that contain significant posture changes need to be transmitted, while those without significant changes are identified~\cite{Head_Pose_Est} and omitted from the transmission.
In this setting, the receiver can use real-time transmitted audios and effective video clips to continuously generate synchronous lip movements in audiovisual clips at the receiver.
Next, we introduce the audio and video semantic communication model in Wav2Vid, and then present video generation process.


Given an original audiovisual clip $\boldsymbol X$ to be transmitted, we first split the video into audio $\boldsymbol A$ and silent video clip $\boldsymbol V$.
To transmit the data through wireless channels, semantic features of $\boldsymbol A$  and $\boldsymbol V$ are extracted for efficient communication. The process of audio semantic extraction is 
\begin{equation}   
\label{equ:encoder_F_Audio}
\boldsymbol{S}^{\mathrm A}= {f}_{\theta}^{\mathrm{E}}(\boldsymbol{A}),
\end{equation} 
where  ${f}_{{\theta}}^{\mathrm{E}}(\cdot)$ indicates the function of the audio encoder parameterized by $\theta$, and $\boldsymbol{S}^{\mathrm A}$ refines the waveform features of audios. 
The video semantic extraction includes contextual feature coding, which is given by
\begin{equation}   
\label{equ:encoder_F_Video}
\boldsymbol{S}^{\mathrm V}= {f}_{{\phi}}^{\mathrm{E}}(\boldsymbol{V}),
\end{equation} 
where${f}_{\phi}^{\mathrm{E}}(\cdot)$ is the video semantic encoder function and $\boldsymbol{S}^{\mathrm V}$ is the video semantic features that contain video spatial-temporal contextual information.
For synchronization, the semantic features of audio and silent video clips are selectively integrated into one data stream $\boldsymbol{S}$, which is given by
\begin{equation} \label{equ:X_hat}
\boldsymbol{S} = \begin{cases} \{\boldsymbol{S}^{\mathrm A}, \boldsymbol{S}^{\mathrm V}\}, \ \ \ \ \ 
  \mathrm{HdPoseEst}(\boldsymbol{V}) \geq \epsilon, \\
\{\boldsymbol{S}^{\mathrm A} \},     \ \ \ \ \ \   \ \ \ \ 
\mathrm{HdPoseEst}(\boldsymbol{V}) < \epsilon , \\
\end{cases}
\end{equation}
where $\epsilon$ is the threshold, and {\color{black} $\mathrm{HdPoseEst}(\cdot) = |\Delta l_{yaw}(\boldsymbol{V})| + | \Delta l_{pitch} (\boldsymbol{V})|+ | \Delta l_{roll} (\boldsymbol{V}) |$  
is the head pose estimation function that
calculates the change in head orientation~\cite{Head_Pose_Est}, with $\Delta l_{yaw}(\boldsymbol{V}), \Delta l_{pitch} (\boldsymbol{V})$, and $\Delta l_{roll} (\boldsymbol{V})$ being the changes of the yaw, pitch, and roll angles of the head in video $\boldsymbol{V}$, respectively.
Thus, $\mathrm{HdPoseEst}(\boldsymbol{V}) \geq \epsilon$ indicates that the speaker's head pose in $\boldsymbol{V}$ changes significantly. }


The process of the data stream transmitting through the wireless channel is 
\begin{equation}
\label{equ:channel}
\widehat{\boldsymbol{S}}={\color{black} h}  \boldsymbol{S}+\boldsymbol{n},
\end{equation}
where $\widehat{\boldsymbol{S}}$ is the received semantic information at the decoder with transmission impairments, $h$ is Rayleigh channel coefficient~\cite{Xing_WCL}, 
and $\boldsymbol{n} \sim \mathcal{N}\left(0, \sigma^{2} \boldsymbol{I}\right)$ denotes Gaussian channel noise with $\sigma^{2}$ being noise variance and  $\boldsymbol{I}$ being identity matrix.

The receiver at the audience uses semantic decoders (audio and video decoders) to decode the received semantics $\widehat{\boldsymbol{S}} = \{ \widehat{\boldsymbol{S}}^{\mathrm A}, \widehat{\boldsymbol{S}}^{\mathrm V} \}$ into reconstructed audio $\widehat{\boldsymbol A}$ and video $\widecheck{\boldsymbol V}$.
The audio decoding process is 
$\widehat{\boldsymbol{A} }
= {f}_{\theta}^{\mathrm{D}}(\widehat{\boldsymbol{S}}^{\mathrm A} )$,  
where ${f}_{\theta}^{\mathrm{D}}$ is the audio decoder function parameterized by $\theta$.
The video decoding process is
$ \widecheck{\boldsymbol V}
= {f}_{ \phi}^{\mathrm{D}}(\widehat{\boldsymbol{S}}^{\mathrm V}) $,
where ${f}_{\phi}^{\mathrm{D}}$ is the video decoder function parameterized by $\phi$.
Note that, when only audio semantics are transmitted and accordingly $\widehat{\boldsymbol S }= 
\{ \widehat{\boldsymbol S}^{\mathrm{A} } \}$, 
the receiver will put the cached $\widecheck{\boldsymbol V}$~(decoded from the latest received $\widehat{\boldsymbol{S}}^{\mathrm V}$) into the video generator.


\begin{figure*}[!ht]
    \centering
    \vspace{-0.1cm}
\includegraphics[width=0.75\linewidth,
    ]{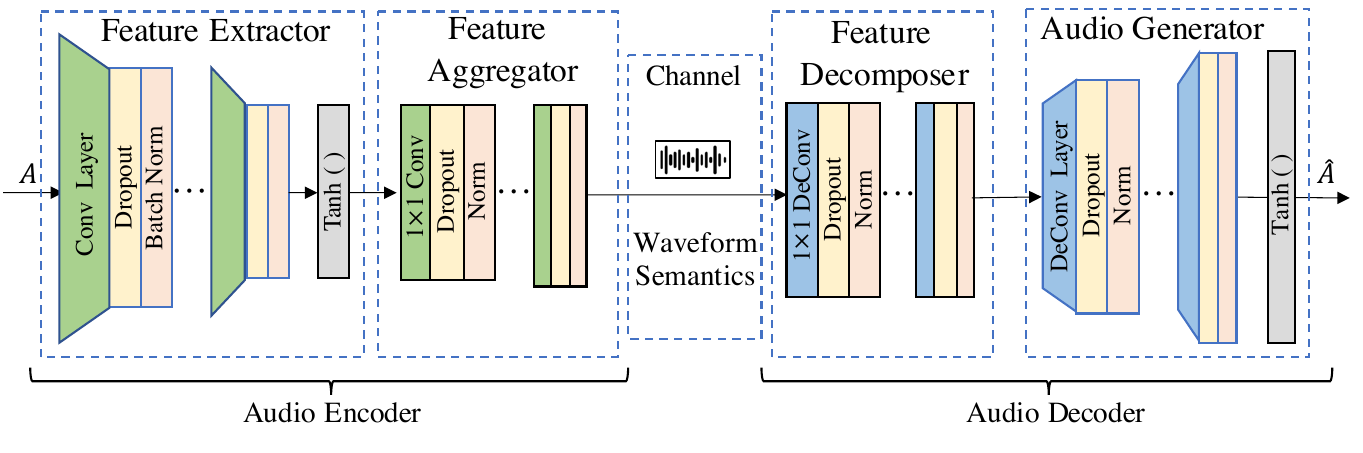}
    \setlength{\abovecaptionskip}{-0.3cm}
    \caption{The architecture of ASC based audio codec.
    }\label{fig:ASC_codec}
    \vspace{-0.2cm}
\end{figure*}


The vacant video duration due to untransmitted videos are filled in by video generated at the receiver. 
The video generator generates the video $\widehat { \boldsymbol V}$ with the decoded video clip $\widecheck{\boldsymbol V}$
and decoded audio $\widehat {\boldsymbol A}$, which is given by  
\begin{equation}   
\label{equ:Gen}
\widehat{\boldsymbol X} = \{ \widehat{\boldsymbol V}, \widehat{ \boldsymbol A} \}
= {g}_{\psi}( \widecheck{\boldsymbol V}, \widehat{\boldsymbol A} ),
\end{equation} 
where ${g}_{\psi} (\cdot)$ is video generation function parameterized by $\psi$.
For the video generation, 
{Frechet Inception Distance} (FID) is used to measure generated video frame's perceptual quality~\cite{FID}, given by 
\begin{equation}
\operatorname{FID}( \boldsymbol V, \widehat{ \boldsymbol{V}} )
=\left\|\mu_{\boldsymbol r}-\mu_{{ \boldsymbol{g} }}\right\|^2+\operatorname{Tr}\left(\Sigma_{\boldsymbol r}+\Sigma_{{ \boldsymbol{g} }}-2\left(\Sigma_{\boldsymbol r} \Sigma_{{ \boldsymbol{g}} }\right)^{\frac{1}{2}}\right), 
\end{equation}
where $\boldsymbol{r} = \mathrm{Inception}(\boldsymbol{V})$ is the feature vector of original real video extracted by InceptionV3 model~\cite{FID}, $\boldsymbol{g} = \mathrm{Inception}(\widehat{\boldsymbol{V}})$ is the extracted visual feature vector from the generated video,
$\mu$ is the mean value of feature vector, $\Sigma$ is the covariance matrix, and  $\operatorname{Tr}(\cdot)$ is the trace of the matrix. 
Therefore, the goal of the Wav2Vid system is to generate video with high perceptual quality, formulated as
\begin{equation}
\min_{\theta, \phi, \psi} \quad \operatorname{FID} (\boldsymbol V, \widehat{ \boldsymbol{V}}).
\end{equation}
Due to video's randomness, we design NN based modules to minimize FID of generative video over wireless networks.



\section{Designed Semantic Codec}

In this section, we illustrate our designed semantic codec models and the corresponding training scheme.
The Wav2Vid system consists of following modules: 
1) audio codec,
2) video codec,
and 3）video generator. 


\vspace{-0.5cm}
\subsection{Audio Codec}

The audio codec uses the autoencoder based architecture according to the audio semantic communication~(ASC) codec in~\cite{thn_ASC}, which achieves JSCC for audio transmission, as shown in Fig.~\ref{fig:ASC_codec}.
Since the Wav2Vid system requires high-fidelity audio transmission {\color{black} for both speech and music, the audio codec is designed to extract audio waveform features and ensure their high-precision transmission~\cite{zhijinQin_wave}, rather than using merely speech-oriented semantic communication method (e.g., speech recognition based method in~\cite{speech_reog_SC}).}
In Fig.~\ref{fig:ASC_codec}, the feature extractor in the encoder uses convolution~(Conv) neural network~(CNN) layers to extract the features of the audio waveform. 
The feature aggregator uses $1\times 1$ Conv layer to aggregate the audio features and overcome channel impairments~\cite{thn_ASC}. 
The decoder at the receiver uses deconvolution~(DeConv) neural network~(DeCNN) layers to decompose the features and generate the audio, which is the reverse operation of the encoder.
Since the ASC codec aims to recover the original audio with high fidelity, the loss function of the audio codec is the normalized root mean square error~(NRMSE), given by 
\begin{equation}
\label{equ:NRMSE}
\mathcal{L}_{\theta}
\left( \boldsymbol{A}, \widehat{\boldsymbol{A}}\right) =\\  
\text{NRMSE}(\boldsymbol{A}, \widehat{\boldsymbol{A}}
) = 
\frac{ \sum_{t=1}^{T_a}\left(a_{t}-\widehat{a}_{t}\right)^{2}}{\sum_{t=1}^{T_a} a_{t}^{2}},
\end{equation}
where $\boldsymbol{A} = \left[a_1, \cdots, a_{T_a} \right]$ with $a_t$ being the audio element in $\boldsymbol{A}$ at sampling $t$, $T_a$ is the number of samplings, and   $\widehat{\boldsymbol{A}} = \left[\widehat{a}_1, \cdots, \widehat{a}_{T_a} \right]$ with $\widehat{a}_t$ being the element in $\widehat{\boldsymbol{A}}$ at time $t$. 
The audio codec is trained by end-to-end method~\cite{thn_ASC}.


\vspace{-0.5cm}
\subsection{Video Codec}

\begin{figure*}[!ht]
    \centering
    \vspace{-0.1cm}
\includegraphics[width=0.75\linewidth,
    ]{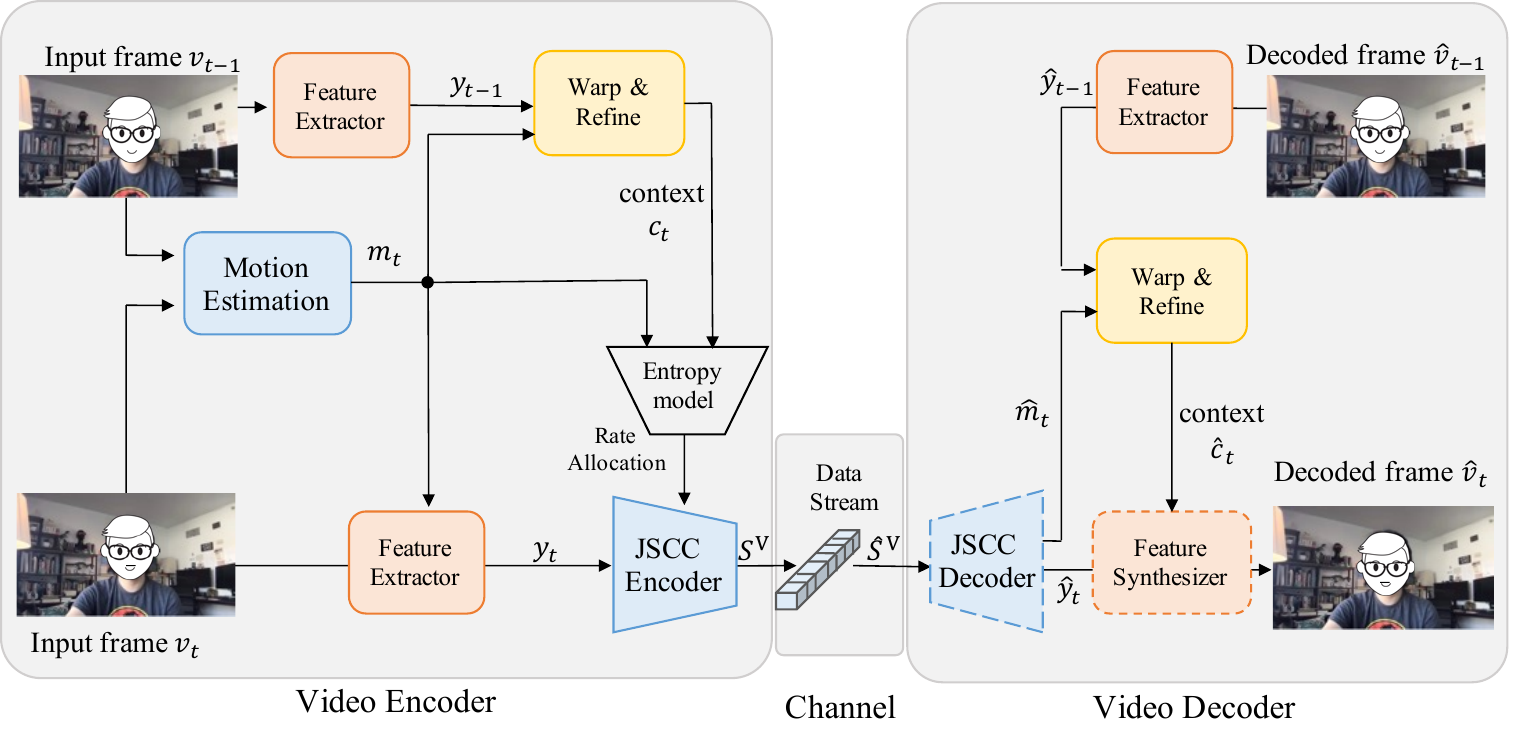}
    \setlength{\abovecaptionskip}{-0.3cm}
    \caption{The architecture of DVST based video codec.
    }\label{fig:Video_codec}
\end{figure*}
\begin{figure*}[!ht]
    \centering
    \vspace{-0.2cm}
\includegraphics[width=0.75\linewidth, 
    ]{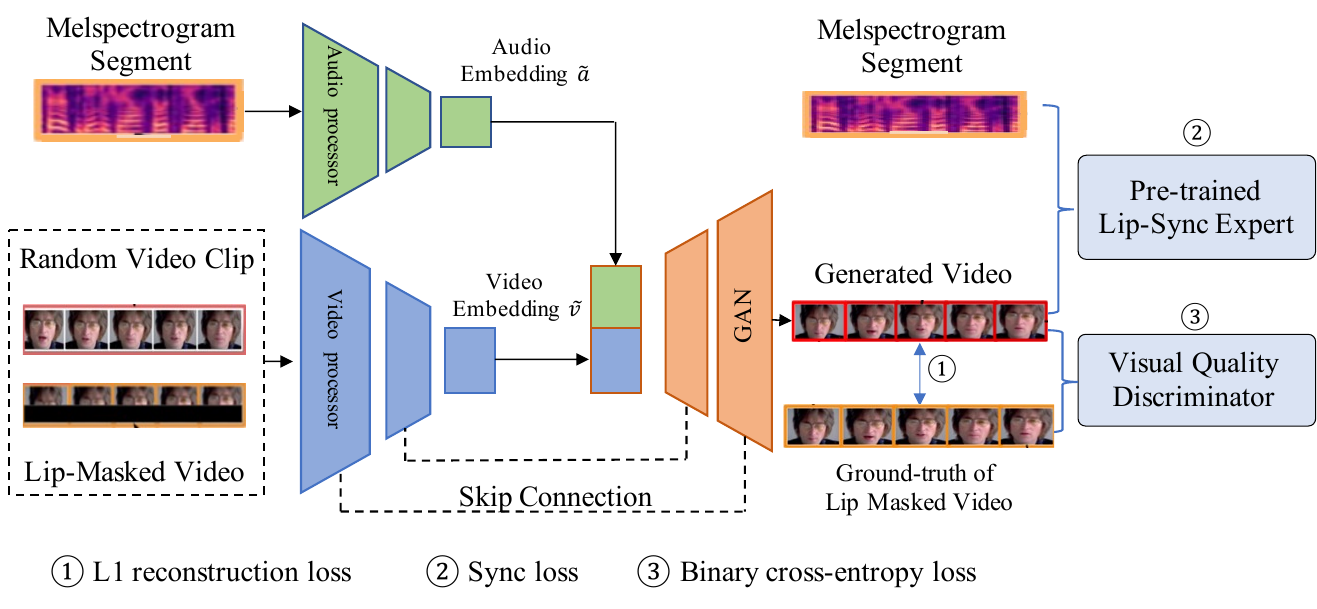}
    \setlength{\abovecaptionskip}{-0.3cm}
    \caption{{\color{black}The architecture of wav2lip video generator.}
    }\label{fig:Video_gen}
    \vspace{-0.3cm}
\end{figure*}

The video codec uses the NN based wireless deep video semantic transmission~(DVST) system according to~\cite{Deep_V_SC_ZP}, as shown in Fig.~\ref{fig:Video_codec}. 
The video codec consists of 2 modules: semantic extraction module and contextual codec module.

\subsubsection{Semantic extraction}
The video codec extracts visual object information into a high dimensional representation using NNs, as shown in Fig.~\ref{fig:Video_codec}. 
In particular, 
the feature extractor processes the frame into a semantic feature map $\boldsymbol{y}_t$.
The motion estimation module obtains the motion vector $\boldsymbol{m}_{t}$ between the reference frame $\boldsymbol{v}_{t-1}$ and the current frame $\boldsymbol{v}_{t}$.
Then, $\boldsymbol{y}_t$ and $\boldsymbol{m}_{t}$ are concatenated, warped (align motion vectors on feature map spatially), and refined (correct spatial discontinuities) in the video context $\boldsymbol{c}_t$. 
In this way, the context $\boldsymbol{c}_t$ contains contextual information between adjacent video frames in a high-dimensional feature space~\cite{DCVC}.
The detailed architectures of feature extractor, motion estimation, and warp \& refine modules are described in~\cite{Deep_V_SC_ZP}.
To compress the video, $\boldsymbol{y}_t$ and $\boldsymbol{m}_{t}$ are encoded into one data stream to be transmitted, while $\boldsymbol{c}_t$ is used to determine the data stream length.

\subsubsection{Contextual encoding}
The contextual codec module
uses an entropy model to determine the data stream length for transmitting coded feature $\boldsymbol{y}_t$ and motion $\boldsymbol{m}_t$~\cite{li2024video}. 
The entropy model fuses the outputs of the temporal prior encoder~(temporal correlation extraction), auto regressive network~(spatial correlation extraction), and hyper prior codec~(hierarchical prior side information extraction)~\cite{Deep_V_SC_ZP}.
The data stream length is increased along with the entropy of $\boldsymbol{y}_t$ and $\boldsymbol{m}_t$, given by
\begin{equation} \label{k}
\begin{aligned}
k_{t} = k_{t}^\mathrm{y} + k_{t}^\mathrm{m} = &  - \sum_{i} \eta_{t,i} \log P \left(\bar{y}_{t, i} \mid {\mathbf{c}}_t, \overline{\mathbf{z}}_t^{\mathrm{y}}, \overline{\mathbf{y}}_{t,<i}\right) \\
& - \sum_{j}  \eta_{t, j} \log P \left(\bar{m}_{t, j}^{} \mid \overline{\mathbf{z}}_t^{\mathrm{m}}, \overline{\mathbf{m}}_{t,<j}^{\mathrm{}}\right),
\end{aligned}
\end{equation} 
where $\eta_{t,i}$ and $\eta_{t,j}$ are scaling factors for the number of symbols to be transmitted; $\bar{(\cdot)}$ is the quantization operation; $\overline{\mathbf{z}}_t^{\mathrm{y}}$ is the quantized hyper prior of $\boldsymbol{y}_t$;  $\overline{\mathbf{y}}_{t,<i}$ is the tensor consisting of quantized feature map $\overline{{y}}_{t,l}$ with $l<i$. 
$\overline{\mathbf{z}}_t^{\mathrm{m}}$ is the quantized hyper prior of $\boldsymbol{m}_t$; $\overline{\mathbf{m}}_{t,<j}$ is the tensor consisting of quantized motion vector $\overline{{m}}_{t,l}$ with $l<j$.
Given $k_t$, the transmitted data stream is encoded by a JSCC codec, according to the estimation feedback of wireless channel conditions~\cite{Deep_V_SC_ZP}.





\textit{Contextual decoding and semantic synthesizer:}
At the receiver, the JSCC decoder recovers the semantic feature map $\widehat{\boldsymbol{y}}_t $ and the motion vector $\widehat{\boldsymbol{m}}_t $, then fused with $\widehat{\boldsymbol{y}}_{t-1}$ to reconstruct the current video frame $\widehat{\boldsymbol v}_{t}$.
The JSCC decoder and the feature synthesizer are the inverse functions of the JSCC encoder and feature extractor, respectively~\cite{Deep_V_SC_ZP}.  
The video codec aims to balance data stream length and peak signal-to-noise ratio~(PSNR) between  $\widecheck{\boldsymbol{V}}$ and $\boldsymbol{V}$, given by  
\begin{equation}
\begin{aligned}
\mathcal{L}_{\phi}\left(\boldsymbol{V},\widecheck{\boldsymbol{V}} \right) &  = \lambda ~ \text{PSNR}  ({\boldsymbol{V}}, {\widecheck{\boldsymbol{V}}}) +\mathrm{len}(\boldsymbol{S}^{\mathrm{V}}) \\
& =  \lambda \cdot 10 \cdot \log_{10} \left( \frac{\max(\boldsymbol{V})^2}{\text{MSE}( {\boldsymbol{V}}, {\widecheck{\boldsymbol{V}}}   )} \right) + \sum_{t}^{T_v} k_t, 
\end{aligned}
\end{equation}
where $\lambda$ is the weight of PSNR to balance video codec's accuracy and compression rate, $\text{MSE}(\cdot, \cdot)$ is the mean square error function,  $\max(\boldsymbol{V})$ is the maximum value of $\boldsymbol{V}$, and $ \mathrm{len}(\boldsymbol{S}^{\mathrm{V}}) = \sum_{t}^{T_v} k_t$ is the length of data stream $\boldsymbol{S}^{\mathrm{V}}$ with $T_v$ being the total frame number in $\boldsymbol{V}$.

\vspace{-0.5cm}
\subsection{Video Generator}

To fill in the vacant video duration caused by the untransmitted video, a video generator is introduced to generate lip movements in receiver's audiovisual video clips $\widehat{\boldsymbol{X}}$, as shown in Fig.~\ref{fig:Video_gen}.
{\color{black}
The video generator is deployed by generative adversarial network (GAN) based wav2lip model~\cite{Wav2Lip} rather than diffusion models~\cite{sora} for affordable complexity by mobile device.}
The video generator aims to align the audio and video features and then to use a GAN to generate lip movements that are synchronized to the speaker's audio.
To achieve this, audio and video {\color{black} embeddings} are first extracted by audio and video processors, respectively, then concatenated and finally input to the GAN. The loss function of the video generator is 
\begin{equation}
\mathcal L_{\psi }=\left(1-w_s-w_g\right) \cdot \mathcal L_{\text {recon }} +w_s \cdot E_{\text {sync }}+w_g \cdot \mathcal L_{g e n},
\label{equ:wav2lip_loss}
\end{equation} 
where $w_s$ and $w_g$ are weights, {
$\mathcal L_{\text {recon }}=\frac{1}{T} \sum_{t=1}^T \left\|  v_t-  \widehat{v}_t\right\|_1 $ is the reconstruction error of $T$ video frames. }
The synchronization loss is 
$E_{\text {sync }}=\frac{1}{T} \sum_{t=1}^T-\log \left(P_{\text {sync}, t}\right), $  
where 
$P_{\text {sync},t}=\frac{  \widetilde v_t \cdot   \widetilde a_t}{\max \left(\|  \widetilde v_t\|_2 \cdot\|  \widetilde a_t\|_2, {\color{black}\kappa}\right)}$
is the synchronization probability of each video frame and audio fragment, with $\widetilde v_t$ being video embedding at time $t$, $ \widetilde a_t$ being audio embedding at time $t$, and {\color{black}$\kappa$ being the noise variable, respectively. $P_{\text {sync},t}$ measures the similarity of audio and video in a high-dimensional feature space.} The loss function of the GAN is  
$\mathcal L_{g e n}=\mathbb{E}_{\widehat v \sim \widehat{\boldsymbol{V}}}[\log (1-D( \widehat v)]$,  
where  $D(\cdot)$ is a binary discriminator that identifies whether a video frame is real ($D(\cdot)=1$) or not ($D(\cdot)=0$).
In (\ref{equ:wav2lip_loss}), $\mathcal L_{\text {recon }}$ and $\mathcal L_{g e n}$ increase the fidelity of generated video and $ E_{\text {sync}}$ guarantees synchronization of generated video and audio, thus improving fidelity of generated audiovisual clips.

\vspace{-0.4cm}

\subsection{Training of the proposed system}
Due to the diverse properties of audio and video, 
we use pre-trained models in~\cite{thn_ASC,Deep_V_SC_ZP,Wav2Lip} to extract audio and video semantic features.
Moreover, the coding modules for wireless transmission are fine-tuned online. 
The training algorithm of the proposed framework is summarized in Algorithm~\ref{alg:Train_alg}.
In particular, the pre-trained NNs for semantic extraction and signal reconstruction are first deployed on the transmitter and receiver.
The coding modules are then randomly initialized in the audio and video codecs.
Then, to improve resistance to channel impairments, the coding modules in {\color{black}$\theta$ and $\phi$} perform a few fine-tunings until convergence. 
This is on the condition that the transmitter and receiver share moderate ground-truth feature data.
{\color{black} Since the features in the video generator $\psi$ do not pass through wireless channel, all NNs in $\psi$ are trained offline by optimizing $\mathcal L_{\psi }$ [9] and do not need fine-tuning online.}
For implementation,
since the audio codec, video codec, and video generator consist of finite CNNs and DeCNN, the complexity of the Wav2Vid system
is affordable~\cite{thn_ASC}, enabling the Wav2Vid system to be implemented on mobile devices.

\begin{algorithm}[t]
	\renewcommand{\algorithmicrequire}{\textbf{Input:}}
	\renewcommand{\algorithmicensure}{\textbf{Output:}}
	\caption{Training algorithm of Wav2Vid system.}
	\label{alg:Train_alg}
	\begin{algorithmic}[1]
	\scriptsize
        \STATE \textbf{initialization:} 
         Deploy the {\color{black}pre-trained modules in $\theta, \phi$, and $\psi$ at the transmitter and receiver~\cite{thn_ASC,Deep_V_SC_ZP,Wav2Lip}.} 
        Initialize modules for JSCC coding in $\theta$ and $\phi$.
	    \STATE \textbf{while} the audio and video codecs do not converge \textbf{do}:  
	   \STATE \quad \textbf{for} each batch of training data \textbf{do}:
          \STATE \quad \quad Input training data batches, transmit $\boldsymbol S$ and obtain $\widehat{\boldsymbol S}$, \\
          \quad \quad calculate $\mathcal{L}_{\theta}\left(\boldsymbol{A}, \widehat{\boldsymbol{A}}\right)$ and $\mathcal{L}_{\phi}(\boldsymbol V, \widecheck{\boldsymbol V}).$
          \STATE \quad \quad Fine-tune the feature aggregator and decomposer in \\ 
          \quad \quad  audio codec, with other NNs being frozen \\
          \quad \quad ${\theta} \leftarrow {\theta}-\eta \nabla_{{\theta}} \mathcal{L}_{\theta}\left( \boldsymbol{A}, \widehat{\boldsymbol{A}}\right).$
          \STATE \quad \quad Fine-tune the contextual JSCC codec in video codec, \\
          \quad \quad with other NNs being frozen \\
        \quad \quad ${\phi} \leftarrow {\phi}-\eta \nabla_{{\phi}} \mathcal{L}_{\phi}\left(\boldsymbol{V},\widecheck{\boldsymbol{V}} \right)$.

	   \STATE \quad \textbf{end for}
        \STATE \textbf{end while} 
    	\end{algorithmic}  
    	\vspace{-0.1cm}
\end{algorithm}
\vspace{-0.4cm}


\

 \vspace{-0.3cm}

\section{Simulation Results and Discussion}

\begin{table}[t]
\vspace{-0.3cm}
\centering
    \caption{NN Parameters.}
        \begin{tabular}{|c|c|c| } 
        \hline 
        \textbf{Model} & \textbf{Module}&
        \textbf{Setting}\\
        \hline
        \multirow{2}*{Audio codec~\cite{thn_ASC}} & 
        {Audio Encoder}  & 
        {7 CNN blocks}  \\
        \cline{2-3}
        ~ & Audio Decoder & 8 DeCNN blocks  \\
        \hline

       \multirow{4}{*}{Video codec~\cite{Deep_V_SC_ZP}} & 
        Feature Extractor & 1 CNN \\
        \cline{2-3}
        & Entropy model & 4 CNNs \\
        \cline{2-3}
        & \multirow{2}{*}{JSCC codec} & 1 CNN + 4 Swin \\
        & & Transformer blocks \\
        \hline
        
        \multirow{2}*{\makecell{Video \\ generator~\cite{Wav2Lip}}} & 
        \makecell{Audio \& video processor}  & 
        {2 CNNs} \\
        \cline{2-3}
        ~ & Generator & 1 DeCNN based GAN  \\
        \hline
        \end{tabular}
    
    \label{tab:para}
    \vspace{-0.4cm}
\end{table}

For the simulation, we consider a video conferencing system that contains one transmitter and one receiver. 
We use real-world audio dataset 
LibriSpeech~\cite{Librispeech} and talking head videos in~\cite{JSAC_mit}.
The Wav2Vid system is trained in Rayleigh channel with varied SNR from 0 to 20 dB.
The semantic extraction NN parameters are set as in~\cite{thn_ASC,Deep_V_SC_ZP},
the NN parameters of the video generator are the same as in~\cite{Wav2Lip}.
The NN parameters for coding in ASC codec and video coedc are listed in Table~\ref{tab:para}.
We compare our proposed method with the following baselines:
{\color{black}
1) Traditional codecs: transmit audiovisual clips using pulse code modulation (PCM)/H.265 + LDPC + 16 quadrature amplitude modulation (QAM) schemes~\cite{thn_ASC};
2) DVST: reconstructs video with NN encoding and decoding video semantics~\cite{Deep_V_SC_ZP};      
3) Txt2Vid: generates videos from text~\cite{JSAC_mit}, using (text and video) semantic codecs + text-to-speech (TTS) + video generator. }
{\color{black} All simulations were performed using a single NVIDIA RTX 4090 GPU.}
The demos of the generated videos are uploaded to url: \url{https://github.com/wcsnSC/Generatedvideos}

\begin{table}[t]
\centering
\caption{Amount of Transmitted Data for One Video Content}
\begin{tabular}{|c|c|c|c|c|}
\hline
~ & Traditional & DVST & Txt2Vid & Wav2Vid \\
~ & / Byte & / symbol & / symbol & / symbol \\
\hline
Video & 15 M & 10 M & {\color{black}2.58 M}  & {\color{black} 2 M}  \\
\hline
Audio  & 1 M & 1 M & - & {\color{black} 600 K} \\
\hline
 Text & - & - & {\color{black} 20 K} & - \\
\hline
\makecell{Total  Amount} & 16 M & 11 M & {\color{black} 2.6 M} & {\color{black} 2.6 M} \\
\hline
\end{tabular} \label{tab:data_amount}
\vspace{-0.3cm}
\end{table}

Table~\ref{tab:data_amount} shows the amount of transmitted data~(byte for the traditional method and symbol for the others) when transmitting one video content with 18s duration. 
{\color{black} The computation time of video generation is nearly 15s, which meets the requirements of real-time generation.}
{\color{black}For fair comparison, the Txt2Vid method transmits longer duration video clip than Wav2Vid, thus the Wav2Vid costs the same bandwidth as Txt2Vid.}
From Table~\ref{tab:data_amount}, we see that the Wav2Vid method can reduce the amount of transmitted data by up to 83.75\%, compared to traditional methods.
This reduction is because Wav2Vid extracts the correlations between videos and audios. 

\begin{figure}[t]
\vspace{-0.2cm}
    \centering
    \includegraphics[width=0.9\linewidth]{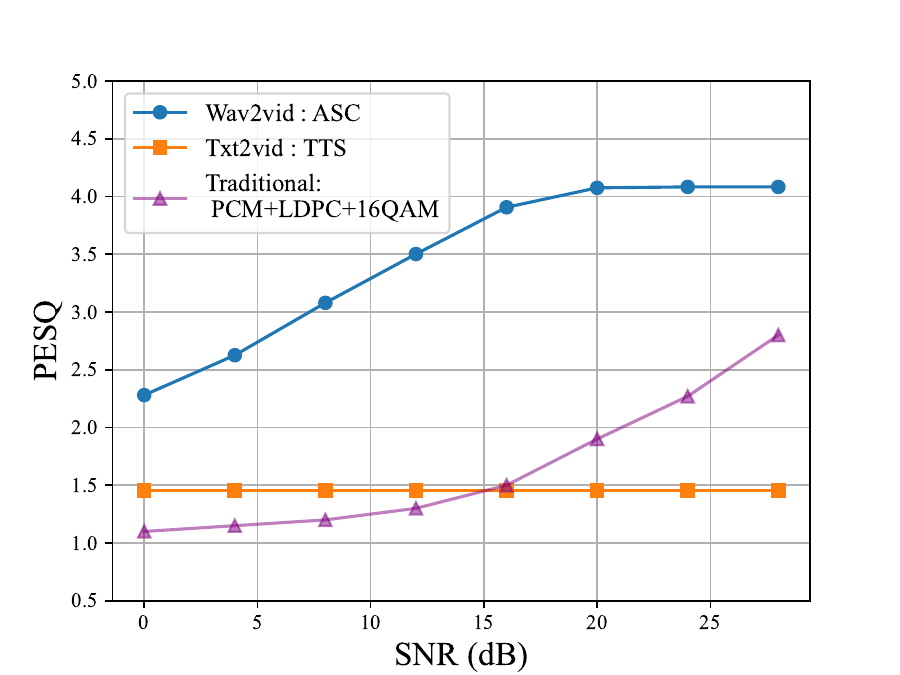}
    \setlength{\abovecaptionskip}{-0.cm}
    \caption{\color{black} Audio PESQ vs SNR.}
    \label{fig:pesq}
    \vspace{-0.4cm}
\end{figure}

\begin{figure}[t]
\vspace{-0.2cm}
    \centering
    \includegraphics[width=0.9\linewidth]{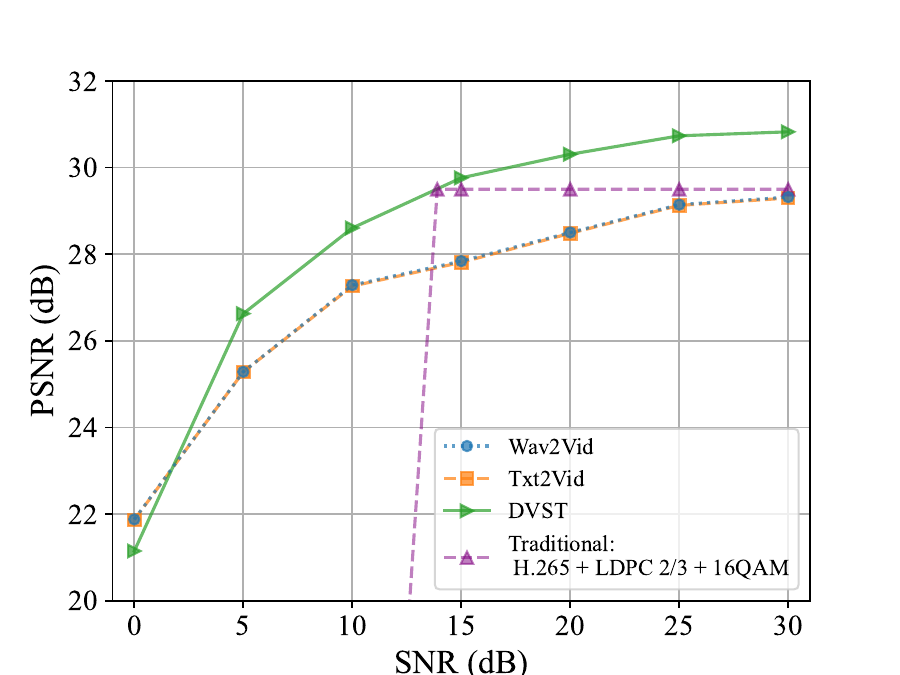}
    \setlength{\abovecaptionskip}{-0.cm}
    \caption{Video PSNR vs SNR.}
    \label{fig:psnr}
\end{figure}

Fig.~\ref{fig:pesq} shows how the audio perceptual evaluation of speech quality (PESQ) of all methods changes as the Rayleigh channel SNR increases. 
 Fig.~\ref{fig:pesq} shows that the ASC based audio codec can improve the audio perceptual quality especially under low SNR regions. 
The low PESQ of Txt2Vid stems from the text to speech~(TTS) technology, which generates constant speed audio that is asynchronous to the original audio, losing emotion and emphasis information and reducing the PESQ.



In Fig.~\ref{fig:psnr}, we show the change of video PSNR  of all methods as Rayleigh channel SNR increases. 
In Fig.~\ref{fig:psnr}, the DVST achieves the highest PSNR due to the semantic extraction and 
the bend of the traditional method's curve is due to the ``cliff effect"~\cite{Deep_V_SC_ZP}.  
The similarity of Wav2Vid and Txt2Vid methods' PSNR stems from the use of the same video generation model, with primary differences limited to lip movements, which do not significantly impact PSNR.
Moreover, when SNR is above 15~dB, the videos generated by the Wav2Vid and Txt2Vid methods achieve high accuracy, with similar PSNR values to those of the reconstructed videos using traditional method.


\begin{figure}[t]
\vspace{-0.2cm}
    \centering
    \includegraphics[width=0.9\linewidth]{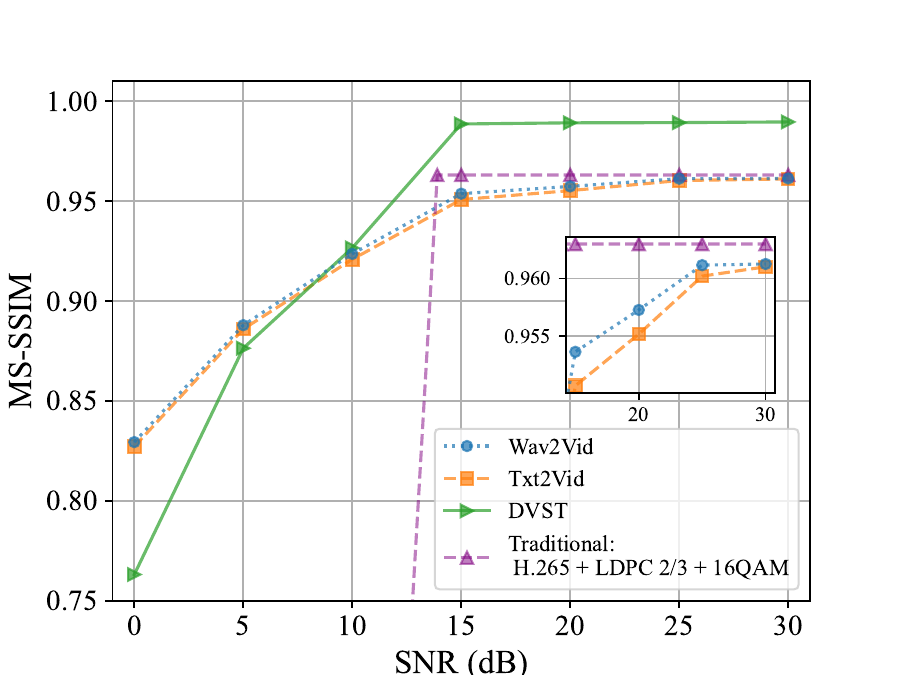}
    \caption{Video MS-SSIM vs SNR.}
    \label{fig:ssim}
\end{figure}

Fig.~\ref{fig:ssim} shows how the multiscale structural similarity (MS-SSIM) of all methods changes as the Rayleigh channel SNR increases.
The results show, when the SNR is higher than 15~dB, the Wav2Vid and Txt2Vid methods can achieve similar MS-SSIM to the traditional method.
When SNR is lower than 8~dB, the MS-SSIM of Wav2Vid and Txt2Vid methods are higher than that of the DVST method, which is because the GAN in video generator is trained with noise, thus achieving more robust performance against noise. 
Fig.~\ref{fig:ssim} also shows that the MS-SSIM of Wav2Vid is slightly higher than Txt2Vid, which stems from the synchronous audio of Wav2Vid method.

\begin{figure}[t]
\vspace{-0.2cm}
    \centering
    \includegraphics[width=0.9\linewidth]{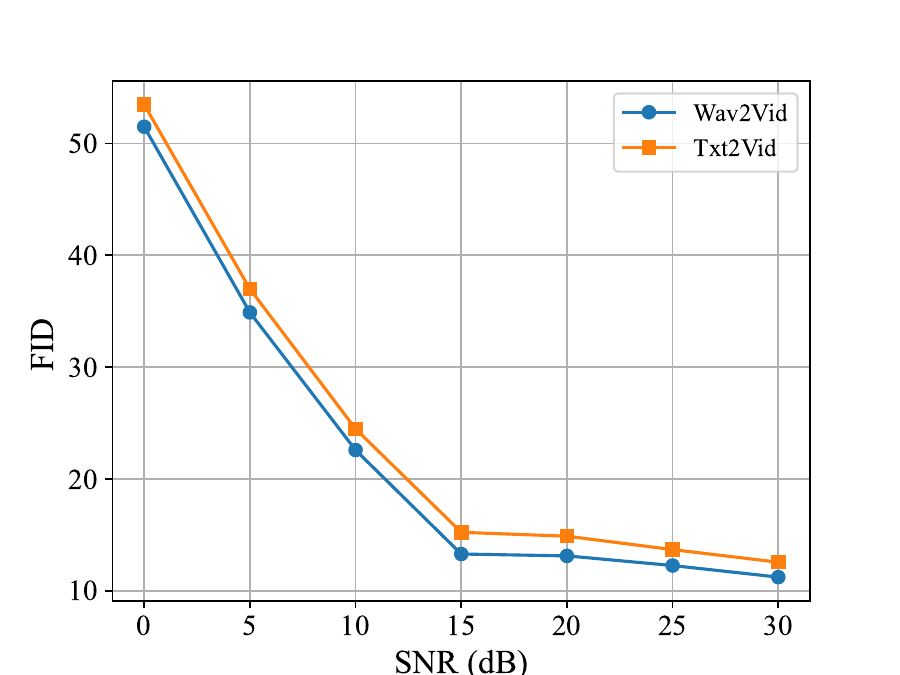}
    \caption{Generative video FID vs SNR.}
    \label{fig:fid}
    \vspace{-0.4cm}
\end{figure}

Fig.~\ref{fig:fid} shows how the FID between the generated video and 
the original video of all Wav2Vid and Txt2Vid changes as Rayleigh channel SNR increases. 
In Fig.~\ref{fig:fid} we see that, the FID of Wav2Vid is lower than that of the Txt2Vid. 
The reason is that 
{\color{black} Wav2Vid can reconstruct the synchronous audio with original video while Txt2Vid generates asynchronous audio generated from text.
Fig.~\ref{fig:fid} demonstrates the video generation improvement of Wav2Vid with transmitting high-fidelity audio rather than text in Txt2Vid.}


\vspace{-0.3cm}

\section{Conclusion}
In this paper, we have developed an efficient multimodal data communication system for video conferencing.
In our system, due to the temporal correlation of video, the transmission of redundant visual video data has been replaced by the audio-driven generative video.
In particular, we have proposed a Wav2Vid system, where the whole duration of audio and short duration video are synchronously transmitted.
The receiver combines the decoded audio and video data, and then uses a GAN to generate the lip movement videos of the speaker.
Simulation results show that the proposed Wav2Vid system can reduce the transmitted data by up to 83\% while maintaining the perceptual quality of the generated video.

\vspace{-0.3cm}
\def\baselinestretch{0.83}

\bibliographystyle{IEEEtran}
\bibliography{IEEEabrv,ref}

\end{CJK}
\end{document}